\documentclass{aastex}
\usepackage{emulateapj5}

\def\gsim{~\rlap{$>$}{\lower 1.0ex\hbox{$\sim$}}}
\def\lsim{~\rlap{$<$}{\lower 1.0ex\hbox{$\sim$}}}

\def\d{{\rm d}}
\def\kms{\,\hbox{km}\,\hbox{s}^{-1}}
\def\ergMsol{\hbox{erg}\,M_{\odot}^{-1}}

\def\erh{\epsilon_{\rm reheat}}
\def\eho{\epsilon_{\rm halo}}
\def\esw{\epsilon_{\rm sw}}

\newenvironment{inlinefigure}{ 
\def\@captype{figure} 
\noindent\begin{minipage}{0.999\linewidth}\begin{center}} 
{\end{center}\end{minipage}\smallskip} 
 
\begin{document}

\title{What Shapes the Luminosity Function of Galaxies?}
\shorttitle{What Shapes the Luminosity Function of Galaxies?}
\author{A.~J.~Benson}
\affil{California Institute of Technology}
\affil{MC 105-24, 1200 E. California Blvd., Pasadena, CA 91125, U.S.A.}
\email{abenson@astro.caltech.edu}
\and
\author{R.~G.~Bower, C.~S.~Frenk, C.~G.~Lacey, C.~M.~Baugh \& S.~Cole, }
\affil{Institute for Computational Cosmology}
\affil{Physics Department, University of Durham}
\affil{South Road, Durham, DH1 3LE, England}
\email{C.M.Baugh,Shaun.Cole,C.S.Frenk,C.G.Lacey@durham.ac.uk}
\shortauthors{Benson et al.}

\maketitle

\begin{abstract}
We investigate the physical mechanisms that shape the luminosity
function of galaxies in hierarchical clustering models.  Beginning
with the mass function of dark matter halos in the $\Lambda$CDM
cosmology, we show, in incremental steps, how gas cooling,
photoionization at high redshift, feedback processes, galaxy merging
and thermal conduction affect the shape of the luminosity function. We
consider three processes whereby supernovae and stellar wind energy
can affect the forming galaxy: (1) the reheating of cold disk gas to
the halo temperature; (2) expansion of the hot, diffuse halo gas; (3)
complete expulsion of cold disk gas from the halo. We demonstrate that
while feedback of form (1) is able to flatten the faint end of the
galaxy luminosity function, this process alone does not produce the
sharp cut-off observed at large luminosities.  Feedback of form (2) is
also unable to solve the problem at the bright end of the luminosity
function. The relative paucity of very bright galaxies can only be
explained if cooling in massive halos is strongly suppressed. This
might happen if thermal conduction near the centres of halos was very
efficient, or if a substantial amount of gas was expelled from halos
by process (3) above. Conduction is a promising mechanism, but an
uncomfortably high efficiency is required to suppress cooling to the
desired level.  If, instead, superwinds are responsible for the lack
of bright galaxies, then the total energy budget required to obtain a
good match to the galaxy luminosity function greatly exceeds the
energy available from supernova explosions. The mechanism is only
viable if the formation of central supermassive black holes and
associated energy generation play a crucial role in limiting the
amount of stars that form in the host galaxy. The models that best
reproduce the galaxy luminosity function, also give reasonable
approximations to the Tully-Fisher relation and the galaxy
autocorrelation function.
\end{abstract}


\section{Introduction}

The luminosity function of galaxies is one of the most basic
properties of the galaxy population; yet it contains many valuable
clues to the process of galaxy formation. The basic physical
mechanisms which determine the form of the luminosity function were
first described by \citet{ro} and \citet{wr}. In this picture, galaxy
formation is regulated by the rate at which gas is able to cool in the
parent dark matter halos. These authors suggested that the sharp
cut-off in the galaxy luminosity function arose from the long cooling
times of gas in high mass halos (or high mass protogalaxies in the
case of Rees \& Ostriker).  The model has been developed by many
authors to follow in great detail the formation of galaxies in a
hierarchical universe. Key improvements are the inclusion of galaxy
merging and the evolution of stellar populations
\citep{wf,cole91,kwg,lacey93,cole94,kauff99,sp,cole00,benson02}. Such
models are now being strongly tested by high-precision measurements of
the galaxy luminosity function from large redshift surveys such as the
2dFGRS and 2MASS \citep{cole2mass,koch01}.

While the key physics of gas cooling and merging are now thought to be
modeled with reasonable accuracy
\citep{benson01,benson02,yoshida02,helly02,voit02}, other physics
crucial to establishing the shape of the luminosity function remain
poorly understood. The first uncertainty is the ``feedback'' needed to
regulate the formation of dwarf galaxies, and hence reconcile the
rather shallow slope of the faint end of the observed luminosity
function with the relatively steep mass function of dark matter halos.
While outflows of gas from galaxies have been observed at both low and
high redshift \citep{martin99,pettini02}, the complex physics at work
has not yet been understood in detail, and most models of galaxy
formation simply adopt phenomenological rules to describe their
effects.  Previous work has typically assumed that the dominant
feedback mechanism is the reheating of cold gas in the disk to the
temperature of the diffuse gas halo \citep{wf,cole94,kauff99,efst00},
although complete expulsion of disk gas from the halo was considered
by \cite{sp}. However, although this form of feedback solves the
problem with the faint-end slope of the luminosity function (as it was
originally designed to do), it creates a second problem matching the
very sharp (exponential with luminosity) cut-off seen at the bright
end.  In addition, the effectiveness with which the gas needs to be
reheated seems excessive compared to observations of galactic outflows
\citep{martin99} and to what is seen in simulations of this
process in individual galaxies \citep{strickland00}. These papers
suggest that the mass in the outflow is comparable to the gas mass
that is turned into stars.

All current models of galaxy formation, calculated using either
gas-dynamical simulations \citep{pearce01,kay02,katz02,weinberg03} or
semi-analytic techniques (e.g. \citealt{kauff99,sp,cole00}), exhibit
strong gas cooling in the central regions of groups and clusters. This
leads to the formation of extremely bright galaxies (which are never
seen in reality) unless some additional suppression of the gas cooling
is assumed.  The suppression mechanisms which have been considered in
semi-analytical models are: simply switching off cooling and/or star
formation in the most massive halos, e.g. \citet{kauff99};
redistributing the gas within the halo so that it becomes so rarefied
that its cooling time is longer than the age of the Universe,
e.g. \citet{cole00}; or suppressing cooling until the halo is
completely virialized, e.g. \citet{vankampen99}.  The model presented
in \citet{cole00} adopted a low value for the cosmic baryon fraction
($\Omega_{\rm b}=0.02$), and a model for the halo gas distribution in
which the core radius was a function of the amount of gas that had
cooled in previous generations of halos. In combination, these
assumptions were able to produce a good match to the galaxy luminosity
function; however, both are now disfavored by current observational
data (e.g. \citealt{omeara,allen01,ettori02}). If the more recent,
higher value of the baryon abundance ($\Omega_{\rm b}
\approx 0.04$) is adopted, the growing core radius has little effect
on the cooling rates (since the cooling radius is then significantly
beyond the typical core radii).

In this paper, we investigate physical processes that may be
responsible for shaping the bright end of the luminosity function.  In
addition to the conventional feedback process, in which 
cold gas in the disk is reheated to the temperature of the diffuse gas halo,
the feedback energy may be used to regulate the formation of the
galaxy in two further ways. While the wind flowing from the gas disk
may contain a relatively small mass, the energy in the disk outflow
may be transferred to the existing hot gas halo, causing it to expand
within the halo potential. This makes the central gas more diffuse,
lengthens its cooling time and so reduces the rate at which gas is
supplied to the cold disk. This type of process is seen in simulations
of the effect of injection of relativistic radio-emitting plasma in
the centres of clusters \citep{quilis01,bruggen02}. An 
alternative is to assume that the gas expelled from the disk does not
mix with the virialized diffuse gas halo, and that the energy per
particle is sufficiently high that the material escapes the confining
potential and is never recaptured.

A further means by which the supply of cold gas to the galaxy disk may
be reduced is suggested by recent Chandra and XMM observations of
galaxy clusters
\citep{peterson,tamura,fabian01,johnstone,bohringer,mcnamara,nulsen}.
These have led to a revival of the idea that thermal conduction may be
an important source of heating in the central regions of clusters
\citep{narayan01,gruzinov,voight}. Indeed, the conductivity has been
inferred to be close to the Spitzer rate expected for a fully ionized
plasma \citep{fabian02}. As first suggested by \citet{fabian02},
heating due to conduction could plausibly counteract the excessive
cooling predicted to occur in the most massive halos and thereby
explain the dearth of highly luminous galaxies in the Universe.

In the remainder of this paper, we will demonstrate quantitatively the
importance of each physical mechanism and its role in setting the form
of the luminosity function. We will demonstrate that the milder forms
of feedback are unable to account simultaneously for the sharp break
in the luminosity function and for the flat faint-end slope.  The two
processes discussed above, thermal conduction and superwinds, may
provide the answer to this problem. Thermal conduction is capable of
suppressing the formation of the brightest objects if the conductive
efficiency is high. Alternatively, a model that includes energetic gas
expulsion is also able to produce reasonable fits to the observed
luminosity function, but the energy required to power this
``superwind'' is larger than that available from supernova explosions
and stellar winds.

\section{The Model}
\label{sec:model}

We compute the luminosity function of galaxies in a cold dark matter
Universe using a development of the semi-analytic model, {\sc
galform}, described by \citet{cole00}. Our changes to the model
concern the treatment of the gas distribution in halos, and the
treatment of feedback.

\subsection{Feedback}
In \citet{cole00}, the star formation rate was affected by feedback
generated when cold gas is reheated by energy injected by supernovae
explosions and stellar winds. This process transfers gas from the
star-forming disk of the galaxy to a diffuse corona or halo. The rate
at which cold gas is reheated is assumed to be related to the star
formation rate, $\dot{M}_{\star}$, by:
\begin{equation}
 \dot{M}_{\rm reheat}= \left(V_{\rm disk}\over V_{\rm
hot}\right)^\alpha \dot{M}_{\star},
\label{eq:reheat_cole}
\end{equation}
where $V_{\rm disk}$ is the circular speed at the half-mass radius of
the gas disk, and $V_{\rm hot}$ and $\alpha$ are adjustable parameters.
Here we include this feedback mechanism, but we also consider two
further schemes which correspond to different processes by which the
energy from supernova explosions and stellar winds can regulate the
gas cooling and star formation rates. A further fraction of the total
energy will be radiated away and so is not available for feedback.  We
assume that the way the energy is apportioned amongst the three
processes is fixed by the local supernova environment, and does not
depend on the mass of the galaxy or on the instantaneous star
formation rate.\footnote{In principle the supernova's local
environment---the interstellar medium---could depend on global
properties of the galaxy, such as its mass, but we neglect any such
dependence here.} Thus, feedback in our model is specified by three
parameters: $\erh, \eho, \esw$. It is useful to note that the total
energy available from supernovae is of order $10^{49} {\hbox{erg}}$
for each Solar mass of stars that is formed, for a standard IMF.

\begin{enumerate}
\item Disk reheating. This is the feedback scheme explored by
\citet{cole00}. Gas is removed from the disk of the galaxy at a rate
proportional to the star formation rate. This gas can be thought of as
either being ejected at close to, but below the escape velocity of the
halo, or as being heated to the halo virial temperature. In the first
case, the gas will subsequently be heated back to the halo virial
temperature, through mixing with the halo gas, or through further
shocks when a new halo forms. In either case, the gas does not leave
the halo and so is available for cooling in the future. We
parameterize the energy invested in reheating the cold disk gas as $\erh\,
10^{49} {\hbox{erg}}$ per Solar mass of stars formed. If we
approximate the gravitational potential wells of galaxies as being
self-similar, then the energy required to eject the gas from the disk
should scale as $V_{\rm disk}^2$, and one obtains
formula~(\ref{eq:reheat_cole}) above
with $\alpha=-2$. Specifically, if we assume that gas is ejected with
energy per unit mass equal to $V_{\rm disk}^2$, then the rate at which gas
is reheated is given by: 
\begin{equation}
  \dot{M}_{\rm reheat}= \frac{5.6 \erh}{\left(V_{\rm disk}/
  300\kms\right)^2} \dot{M}_*.
\end{equation}
We therefore have the
following relation between $\erh$ and $V_{\rm hot}$:
\begin{equation}
   \erh = 0.18 \left(V_{\rm hot}\over 300\kms\right)^2,
\end{equation}
If the circular speed of the disk is equal to $V_{\rm hot}$, the mass
of gas reheated is equal to the mass formed into stars. Feedback of
this form is motivated by the study of \cite{efst00}, although the
efficiency assumed here is somewhat higher.

\item Energy injection into the gaseous halo. The feedback energy is
assumed to redistribute the hot, diffuse gas so that the central
density of the gas is reduced, and the central cooling time is
increased.  The energy injected per unit mass of stars formed is
parameterized as $\eho 10^{49} \ergMsol$. The nature of this type of
feedback process can be understood as follows: the total energy
injected into the halo grows with time until it becomes comparable to
the gravitational binding energy of the gaseous halo. At this point,
the cooling 
rate drops dramatically. Since the binding energy of a
halo is proportional to $M_{\rm halo}^{5/3}$, the fraction of the
total gaseous mass which must be turned into stars to achieve this
balance is strongly dependent on halo mass. Larger halos are therefore
able to cool a larger fraction of their baryon reservoir to form
galaxies. This mechanism is described in more detail in
\citet{bower01} and, as noted above, is motivated by simulations of
the injection of radio plasma in cluster centres
\citep{quilis01,bruggen02}.

\item Gas expulsion from the halo. In this scheme, an energy of
$\esw\,10^{49}\ergMsol$ is invested in heating a small mass of the cold gas
disk to an energy much greater than the binding energy of the halo. If
the energy is contained within this super-heated phase (and is not
shared with the diffuse gas halo, as assumed in mechanism 2 above),
the wind may be able to escape completely from the halo. This
mechanism is motivated by the strong, high velocity winds inferred to exist
around Lyman-break galaxies \citep{pettini02,adelberger02} and has
recently been studied theoretically by \citet{shu03}. We assume that
the energy is injected with a thermal distribution of particle
energies, i.e. so that the fraction of mass ejected with energy in the
range $E$ to $E+\d E$ is proportional to $\exp (-E/E_{\rm av}) \d E$,
where $E_{\rm av}$ is the mean ejection energy per particle. The fraction
of the superwind gas with sufficient energy to escape the halo is
computed using this distribution. Some of the expelled gas will be
recaptured when a deeper potential well forms. To estimate this, we
calculate the fraction of the expelled gas with energy less than the
depth of the new potential well, and allow this fraction to be
recaptured. This process continues as deeper potential wells form
until either all of the gas is recaptured, or $z=0$ is reached.

We assume that the mass flux of material in the wind as it leaves the
disk, $\dot{M}_{\rm superwind}$, is proportional to the star formation
rate, with coefficient of proportionality $\beta_{\rm SW}$. The energy
in the superwind as it leaves the disk is given by
\begin{equation}
\dot{E}_{\rm superwind} = 10^{49} \epsilon_{\rm SW} \dot{M}_{\star}
\hbox{ ergs},
\end{equation}
where $10^{49} \epsilon_{SW}$ ergs is the energy per unit mass of star
formation. The characteristic specific energy of the wind is simply
$E_{\rm av}=\dot{E}_{\rm superwind}/\dot{M}_{\rm superwind} \equiv
\frac{1}{2}V^2_{\rm SW}$, and so the characteristic velocity of the
wind is
\begin{equation}
V_{\rm SW} = 1002 \sqrt{\epsilon_{\rm SW}/\beta_{\rm SW}} \hbox{ km/s}.
\end{equation}
We assume that material ejected from the disk requires an energy per
unit mass $\lambda_{\rm SW} V_{\rm disk}^2$ in order to escape also
from the halo. Therefore, in the case in which all of the wind
material has the same energy, superwinds will be driven from halos
with $V_{\rm disk}<709\sqrt{\epsilon_{\rm SW}/\beta_{\rm
SW}\lambda_{\rm SW}}$ km/s, and not from deeper potential wells. For
typical halos in our model $\lambda_{\rm SW}=2.9$ gives a good
estimate of the energy required to escape the combined gravitational
pull of the galaxy and its dark matter halo, and so we use this value
throughout this work. If we assume a distribution of energies in the
wind, then there will be a smooth transition, as $V_{\rm disk}$
increases, between the regime in which the wind escapes and the regime
in which it is retained. Thus, the actual mass flux in the superwind
escaping from the halo is given by
\begin{equation}
\dot{M}_{\rm superwind} = f_{\rm eject}(E_{\rm av}/\lambda_{\rm
SW}V_{\rm disk}^2) \: \beta_{\rm SW} \dot{M}_{\star}, 
\end{equation}
where, for our chosen energy distribution, $f_{\rm
eject}(x)=\exp(-x)$. The material which does escape may be recaptured
by larger halos forming at later times. Our estimate of the fraction
of the wind which escapes and that which can be recaptured by
larger halos is described in Appendix~\ref{app:fracs}.

We regard this mechanism as being more uncertain than mechanisms 1 and 2
because the expelled gas must leave the halo without sharing its
energy with the diffuse hot component. It must therefore punch a well
collimated hole through the halo, or remain contained within a buoyant
bubble which is convected out of the halo \citep{quilis01,springel02,
kay02}.
\end{enumerate} 

In general, it seems plausible that some fraction of the total
feedback energy will be processed into each one of these forms, and we
consider models which include a combination of these processes on an
equal footing with models that include only one.

\subsection{Halo Gas Distribution}
The hot gaseous component in dark matter halos is assumed initially to
have a density profile, $\rho_{\rm g}(r)$, given by the $\beta$-model,
ie, at radius $r$ in the halo (whose dark matter density profile is
assumed to have the NFW profile, \citealt{nfw96,nfw97}),
\begin{equation}
\rho_{\rm g}(r) = {\rho_0 \over [1 + (r/r_{\rm c})^2]^{3\beta/2}},
\end{equation}
where $\rho_0$ is the density at the centre of the halo, $r_{\rm c}$
is the radius of the ``core'' and $\beta$ is a parameter which sets
the slope of the profile on scales larger than $r_{\rm c}$.

Departing from the prescription of \citet{cole00}, we adopt a gas
density profile in the absence of energy injection with fixed 
$r_{\rm c}=0.07 r_{\rm vir}$ (and 
$\beta=2/3$) in all halos. This provides a reasonable match to
gas-dynamic simulations of non-radiative gas in clusters
(e.g. \citealt{eke98}) and to the observed X-ray profiles of relaxed
clusters (e.g. \citealt{allen01}). The normalization of the profile is
determined by the total diffuse gas mass remaining in the halo, and
the temperature of the gas is set assuming hydrostatic equilibrium. As
a boundary condition, we set the temperature at the virial radius
equal to the virial temperature (eqn. 4.1 in
\citealt{cole00}).  This ``default'' profile is modified if the
diffuse gas gains further energy (``excess energy'') as a result of
energy injection (process~2 in Section~2.1). We use the prescription
described in \citet{bower01}, in which the excess energy first causes
the slope of the gas profile to decrease down to a minimum value of
$\beta_{\rm min}= 0.2$, after which it increases the boundary
temperature of the gas halo. The effect of the excess energy is to
decrease the central density of the gas, lengthening its cooling
time. Mass is conserved by pushing some of the diffuse gas outside the
halo; however, this gas can be recaptured as the total halo mass (and
thus the gravitational binding energy) increases.  We assume that the
excess energy is conserved during mergers between halos, although the
results are not qualitatively affected by a small dilution or
amplification of energy during halo mergers.

We define the effective cooling time at radius $r$, $t_{\rm
cool}^\prime(r)$, as the maximum true cooling time (i.e. that defined
by \citealt{cole00}) occurring at smaller radii, plus the free-fall
time from $r$ to the halo centre. This ensures that $\d t_{\rm
cool}^\prime/\d r > 0$, so the cooling radius is always a smoothly
increasing function of radius. Experiments with different approaches
show that the results we present here are not sensitive to the details
of this prescription.

\subsection{Conduction}

Conduction in the ionized gas can transport energy into the inner
regions of the halo, effectively increasing the cooling time of the
gas there. The rate at which energy is deposited into the shell
between radii $r$ and $r+\d r$ is given by:
\begin{equation}
\Sigma \d r = 4 \pi {\d \over \d r} \left( \kappa r^2 {\d T \over \d
r}\right) \d r,
\end{equation}
where the conductivity, $\kappa$, may depend on radius through its
temperature dependence. We approximate $\Sigma$ as
\begin{equation}
\Sigma = \alpha_{\rm cond} 4 \pi \kappa_{\rm s} T,
\end{equation}
where $\kappa_{\rm s}$ is the Spitzer conductivity \citep{spitzer},
and $\alpha_{\rm cond}$ is a parameter that absorbs the dependence on
the shape of the temperature profile as well as any difference between
the actual conductivity and the Spitzer rate. For a power-law
temperature profile, $T\propto r^a$, and conductivity of the Spitzer
form, $\kappa_{\rm s}\propto T^{5/2}$, $\alpha_{\rm cond}=f_{\rm sp} a
(1+7a/2)$ where $f_{\rm sp}$ is the ratio of the conductivity to the
Spitzer value. Adopting a temperature profile with $a=0.4$ as suggested
by recent X-ray observations \citep{voight} gives $\alpha_{\rm
cond}=0.96 f_{\rm sp}$. Adopting a linear temperature gradient gives
$\alpha_{\rm cond}=4.5 f_{\rm sp}$. 

The heating rate due to conduction is subtracted from the radiative
cooling rate to give a net cooling rate for the gas. This net cooling
rate is used to compute the cooling time. Conduction causes the cooling
radius to become smaller than in the standard model. The result is a
suppression of cooling in hot halos, and the mass at which this effect
becomes important is determined by the parameter $\alpha_{\rm cond}$.

\subsection{Photoionization and Merging}

While we employ the detailed model of galaxy merging developed by
\citet{benson02}, we choose to use a much simpler model of the effects
of reionization than used in that paper, in order to incorporate them
easily into our calculation of the galaxy luminosity function. We
simply assume that galaxy formation is completely suppressed by
reionization in dark matter halos with circular velocities below
$V_{\rm reion}$ after redshift $z_{\rm reion}$. Unless otherwise
stated, we adopt $V_{\rm reion}=50\kms$ and $z_{\rm reion}=6$. With
this choice, this simple model matches the results of the full
calculation of \citet{benson02} quite well.

\section{Results}

Throughout this paper, we use the same parameter values adopted by
\citet{benson02}, with the exception of a larger baryon fraction
corresponding to $\Omega_{\rm b}=0.045$, consistent with constraints
from Big Bang nucleosynthesis \citep{omeara}.  We assume a
$\Lambda$CDM universe with mean matter density $\Omega_0=0.3$,
cosmological constant term $\Omega_\Lambda=0.7$, Hubble
constant\footnote{Here and below $h$ denotes the Hubble constant in
units of 100km s$^{-1}$ Mpc$^{-1}$} $H_0=70$km s$^{-1}$ Mpc$^{-1}$ and
linear fluctuation amplitude on spheres of radius 8h$^{-1}$ Mpc,
$\sigma_8=0.9$. Of these parameters, the uncertainty in the value of
$\sigma_8$ has most impact on the model results.  While several
studies support a value of $\sigma_8 \sim 0.9$ (e.g., \citet{bacon02},
\citet{hoekstra02} [gravitational lensing]; \citet{eke96},
\citet{vianna02} [cluster abundance]; \citet{spergel03} [cosmic
microwave background (CMB)]; \citet{contaldi02} [Sunyaev-Zeldovich
effect]), other recent analyses have suggested lower values, $\sigma_8
\sim 0.7$ (e.g., \citet{peacock02} [large scale structure];
\citet{melchiori02} [CMB]; \citet{jarvis02} [gravitational lensing];
\citet{allen02}, \citet{smith03} [cluster abundance]).  A discussion
of recent results may be found in \citet{wang03}. Except where
specified, we show models based on $\sigma_8=0.9$; however, (as we
shall see) taking the lower value, $\sigma_8=0.7$, considerably eases
the energy budget and/or reduces the conduction efficiency required to
match the galaxy luminosity function. Throughout, we use the halo mass
function derived from N-body simulations by
\citet{jenkins01}\footnote{Note that we use the
\protect\citet{jenkins01} mass function to compute the abundances of
halos at $z=0$, but retain the Press-Schechter approximation when
computing halo merger trees. An improved calculation would use a
self-consistent computation of halo merging histories.}, instead of
the Press-Schechter mass function used by \citet{cole00}.

\begin{inlinefigure} 
\resizebox{\textwidth}{!}{\includegraphics{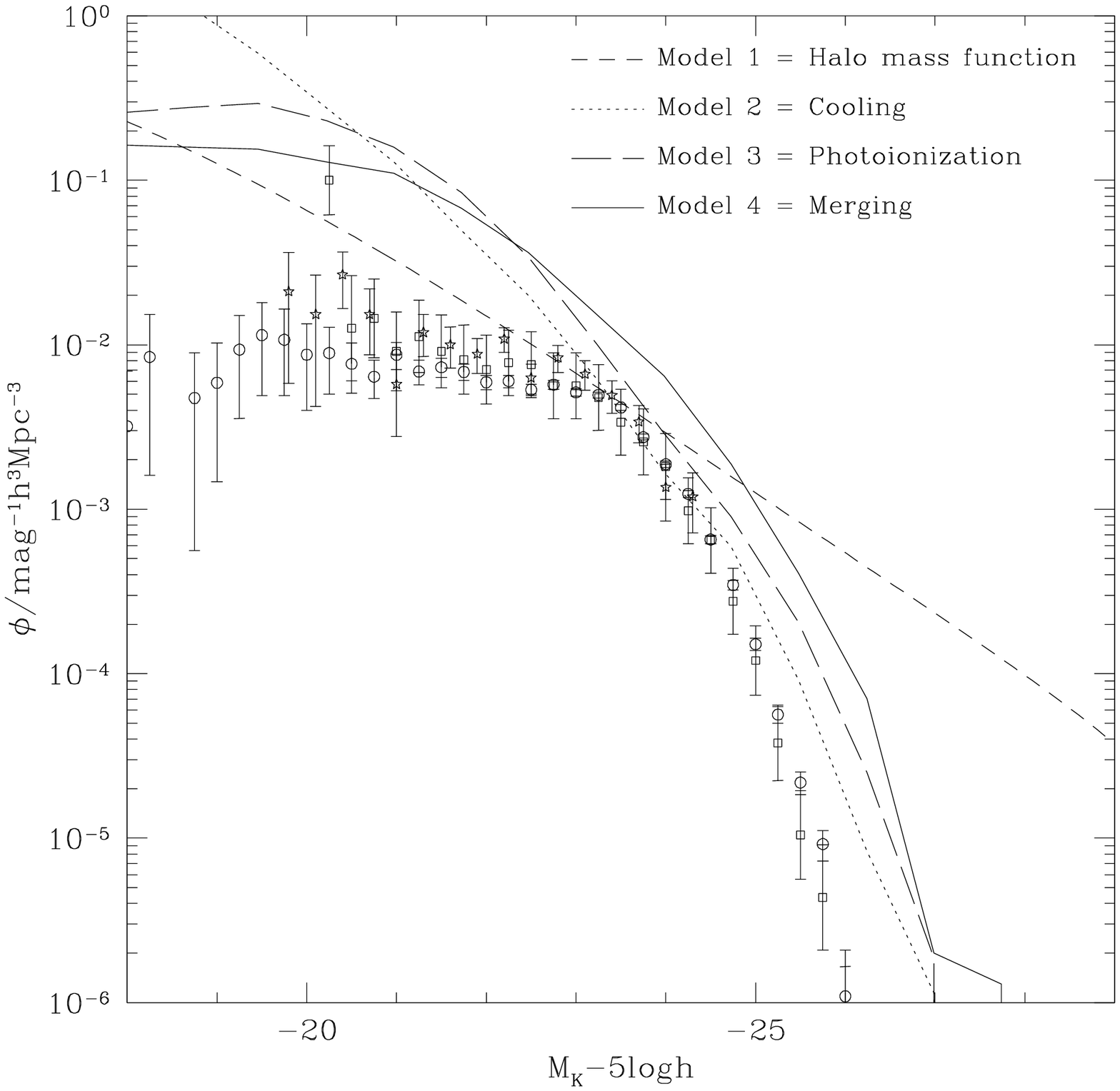}} 
\figcaption{The K-band luminosity function of galaxies. The points show
the observational determinations of
\protect\citet[circles]{cole2mass}, \protect\citet[squares]{koch01}
and \protect\citet[$z<0.1$, stars]{huang02}. Lines show model
results. Model 1 (dashed line) shows the result of converting the dark
matter halo mass function into a galaxy luminosity function by
assuming a fixed mass-to-light ratio chosen to match the knee of the
luminosity function. Model 2 (dotted line) shows the result from {\sc
galform} when no feedback, photoionization suppression, galaxy merging
or conduction are included. Models 3 and 4 (long dashed and solid
lines respectively) show the effects of adding photoionization and
then galaxy merging.
\label{fig:models_123}} 
\end{inlinefigure} 

We compare our model with recent determinations of the K-band
luminosity function assuming a Kennicutt stellar initial mass function
\citep{ken83}. In order to facilitate comparison between models, we
have kept the IMF fixed, and assumed a negligible fraction of brown
dwarf stars\footnote{We set $\Upsilon=1$ in the notation of
\protect\citet{cole00}}. We choose the K-band in order to minimize the
sensitivity of our results to recent star formation and to dust
obscuration. The model of \citet{cole00} includes a detailed and fully
self-consistent calculation of dust extinction which is used in this
work. We find, however, that dust-obscuration has a negligible effect
on our results for the K-band luminosity function (typically shifting
the bright end faintwards by less than one tenth of a magnitude.) For
the observational comparison, we use the local K-band luminosity
functions measured by \citet{cole2mass} and \citet{koch01} (both based
on the 2MASS survey) and the local ($z<0.1$) luminosity function
derived from the much deeper K-band survey of \citet{huang02}. The
analysis by Huang et al. suggests a faint-end slope ($\alpha_{\rm
K}=-1.37\pm0.10$), steeper than the values found by Cole et al
($\alpha_{\rm K}=-0.93$) and by Kochaneck et al. ($\alpha_{\rm
K}=-1.09$). These latter two are also in good agreement with the faint
end slope of the z-band luminosity function measured by Blanton et
al. ($\alpha_{\rm z}=-1.08$) from the SDSS survey. The z-band data
should also be little affected by residual star formation and dust
extinction, but have a deeper surface brightness limit. These
discrepancies indicate that there remain significant systematic
uncertainties in current measurements of the faint end of the K-band
luminosity function, perhaps due to luminosity from low-surface
brightness regions of galaxies being missed as recently suggested by
\citet{andreon}.

\subsection{A Model with constant mass-to-light ratio}

In Fig.~\ref{fig:models_123} we show the simplest possible model of
the luminosity function which we call Model~1 (shown as the dashed
line). In this model, the mass function of dark matter halos
\citep{jenkins01} has been converted into a luminosity function simply by
assuming a fixed mass-to-light ratio ($M/L_{\rm K} = 11 M_\odot/L_{\rm
K,\odot}$), chosen so as to match the knee of the observed
luminosity function. As is well known, this produces a luminosity
function which is much steeper at the faint end than is observed, and
also fails to cut off at bright magnitudes (the halo mass function
does possess a cut-off, but it occurs at much higher mass and lower
abundance than shown in the plot).

\subsection{Model including Cooling only}

\citet{wr} argued that the difference between the halo mass and the
galaxy luminosity functions is due to the dependence of the gas
cooling time on halo mass and to feedback processes.  We use the {\sc
galform} semi-analytic model to follow gas cooling and star formation
in a merging hierarchy of dark matter halos in the $\Lambda$CDM
cosmology. In order to illustrate the simplest possible model first,
we do not include photoionization suppression, feedback, galaxy
merging or conduction. The result is Model~2 (shown as a dotted line
in Fig.~\ref{fig:models_123}). It clearly displays the ``overcooling
problem'': gas has cooled into the smallest halos resolved in the
calculation, producing an overabundance of faint galaxies. As a
result, the faint-end slope of the luminosity function is much too
steep. It should be noted that the results for this cooling-only model
are sensitive to the mass resolution of the {\sc galform} merger
trees. The cooling time for halo gas decreases with decreasing halo
mass down to halo virial temperatures around $10^4$K. Below this
scale, cooling becomes ineffective unless molecular hydrogen is
abundant. This means that for the luminosity function to be fully
converged at all galaxy luminosities, we would need to resolve halos
with virial temperatures as low as $10^4$K at all redshifts at which
there is significant cooling. To obtain the results shown for Model~2,
we ran {\sc galform} at the highest mass resolution that was
computationally feasible, which comes close to resolving $10^4$K halos
at $z=0$. We believe that our results are substantially converged, but
we cannot be certain that they would not change if the mass resolution
were increased further.

\subsection{Model with photoionization and merging}


The problem of mass resolution encountered in Model~2 is effectively
removed once we include feedback processes which are
effective in low-mass halos. Model~3 (long-dashed line in
Fig.~\ref{fig:models_123}) shows the effects of including
photoionization suppression. 
As described by
\citet{benson02}, the formation of low-mass galaxies is suppressed
after the Universe is reionized and, as we noted in \S\ref{sec:model},
we adopt a simplified model of this process which nevertheless is a
good approximation to a full calculation.  The resolution of the {\sc
galform} merger trees is then sufficient to follow the lowest mass
halos that are able to form galaxies, resulting in a converged
solution.  However, photoionization suppression alone is unable to
produce a sufficiently flat faint end in the luminosity function. In
addition, Model~3 also produces more bright galaxies, since less gas
has been locked into stars in the smallest halos.  When both galaxy
merging (which was artificially switched off in Models~1 and~2) and
photoionization are included, as in Model 4 (solid line in
Fig.~\ref{fig:models_123}), the faint end remains too steep and even
more bright galaxies are produced. In the models that follow, we
investigate the effects of including feedback processes in addition to
photoionization and merging.

\begin{inlinefigure} 
\resizebox{\textwidth}{!}{\includegraphics{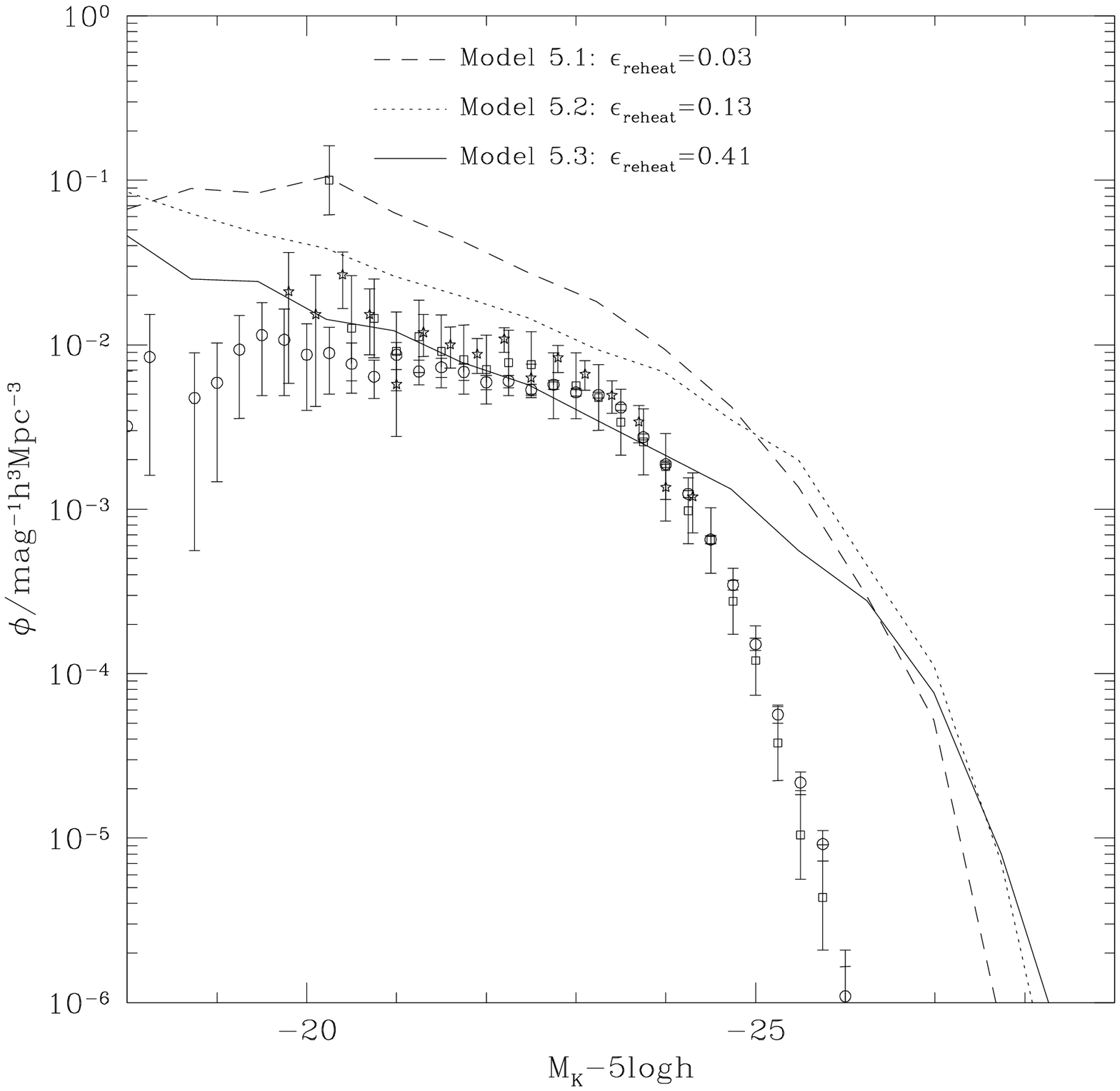}}
\figcaption{Model 5. Starting from Model~4, disk-reheating is added in
order to suppress the formation of small galaxies. Results are shown
for three levels of energy input ($\erh=0.03$, 0.13 and 0.41).  The
data points are the same as Fig.~\ref{fig:models_123}.
\label{fig:models_4}}
\end{inlinefigure} 

\subsection{Model with feedback---disk reheating}

A solution to the faint-end problem is illustrated by Model~5.3 in
Fig.~\ref{fig:models_4}. Here, feedback is included through the
reheating of disk gas in star-forming galaxies. We use the standard
prescription of \citet{cole00}, but with a larger value of $\erh=
0.41$ (equivalent to $V_{\rm hot}=450\kms$) which is required in order
to obtain a similar faint end slope to that in Cole et al.\ for the
larger value of $\Omega_{\rm b}$ assumed in this work. This form of
feedback flattens the luminosity function considerably, resulting in
reasonably good agreement with the observed faint end.  While the
slope is not as flat as that measured by \citet{cole2mass} or
\citet{blanton}, it is in good agreement with the steeper slope
reported by \citet{huang02}. This achievement carries a price,
however---the overabundance of bright galaxies (formed through
excessive cooling in massive halos) is exacerbated, as there is now a
much greater mass of diffuse hot gas remaining in the larger
halos. This gas is sufficiently dense that the central regions are
able to cool; consequently, Model~5.3 produces far too many bright
galaxies. This result depends little on the choice of $\sigma_8$.
Adopting $\sigma_8=0.7$ makes the brightest galaxies only 0.5 mag
fainter.  This clearly demonstrates a longstanding problem in
semi-analytic models: previous calculations have either assumed low
values of $\Omega_{\rm b}$, or have invoked rather artificial ways to
prevent the cooling which forms these overluminous objects.

\begin{inlinefigure} 
\resizebox{\textwidth}{!}{\includegraphics{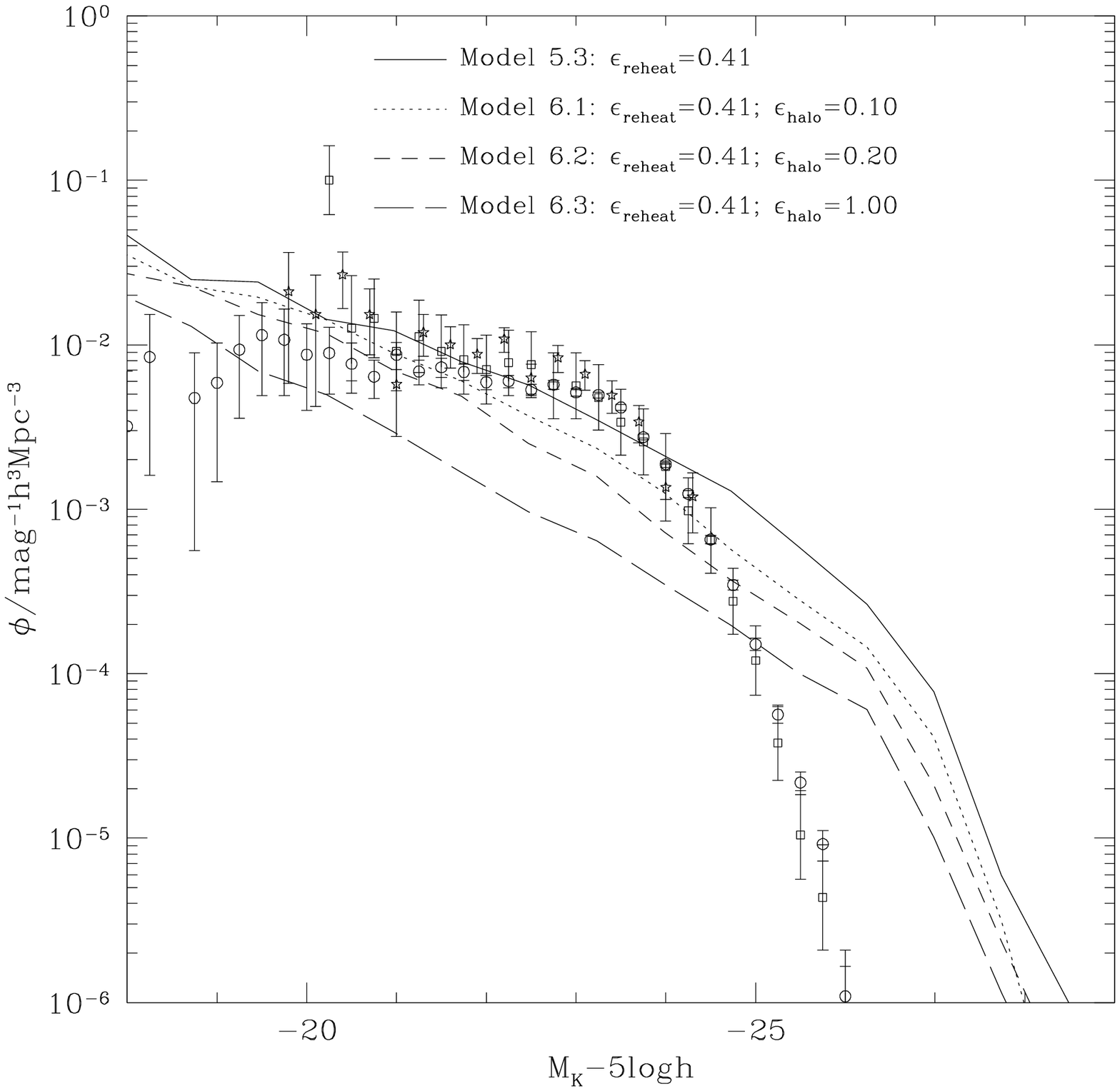}}
\figcaption{Model 6. These models illustrate the effect of energy
injection. Starting from Model~5.3, the effect of heating the diffuse
gas halo is included. Models 6.1--6.3 have $\eho=0.1$, 0.2 and 1.0
respectively, in addition to $\erh=0.41$. Increasing the energy
available for this form of feedback suppresses the formation of both
bright and faint galaxies, but does not produce a sharp break in the
luminosity function.
\label{fig:models_5}}
\end{inlinefigure} 

The energy requirements of Model 5.3 are substantial, but not excessive.
The reheating energy of $0.41 \times 10^{49} \ergMsol$ should be
compared to the total energy available from supernovae explosions
which is approximately $0.7 \times 10^{49}
\,\hbox{erg}\,M_{\odot}^{-1}$ for a Salpeter IMF, or $0.9 \times
10^{49} \,\hbox{erg}\,M_{\odot}^{-1}$ for the Kennicutt IMF adopted in
{\sc galform}.  In a halo of circular velocity 250~km/s, the mass of
gas reheated is more than 3 times the mass of gas formed into
stars. If the level of reheating is reduced, as illustrated by
models~5.1 ($\erh= 0.03$) and~5.2 ($\erh= 0.13$), then the formation
of small galaxies is not suppressed sufficiently to match the observed
luminosity function.

\subsection{Models with energy injection}

In Models~6.1--6.3 (Fig~\ref{fig:models_5}), we investigate the effect
of assuming that a fraction of the supernova and stellar wind energy
heats the diffuse gas halo, causing it to expand. This form of
feedback suppresses the formation of both bright and faint galaxies,
but it does not produce a sharp break in the luminosity
function. Models~6.1 and 6.2 illustrate how as the energy spent in
heating the diffuse halo increases, the break in the luminosity
function becomes less pronounced. If the heating is made even
stronger, the resulting luminosity function approaches a power-law
(Model~ 6.3). This form of feedback clearly cannot solve the problem
of overproduction of bright galaxies.

\subsection{Model with conduction}

We now consider two possible schemes that are capable of producing a
good match to the observed luminosity function. The first involves
balancing radiative cooling with thermal conduction. The second
involves expelling gas from dark matter halos at such high energies
that it is subsequently unable to cool.

\begin{inlinefigure}
\resizebox{\textwidth}{!}{\includegraphics{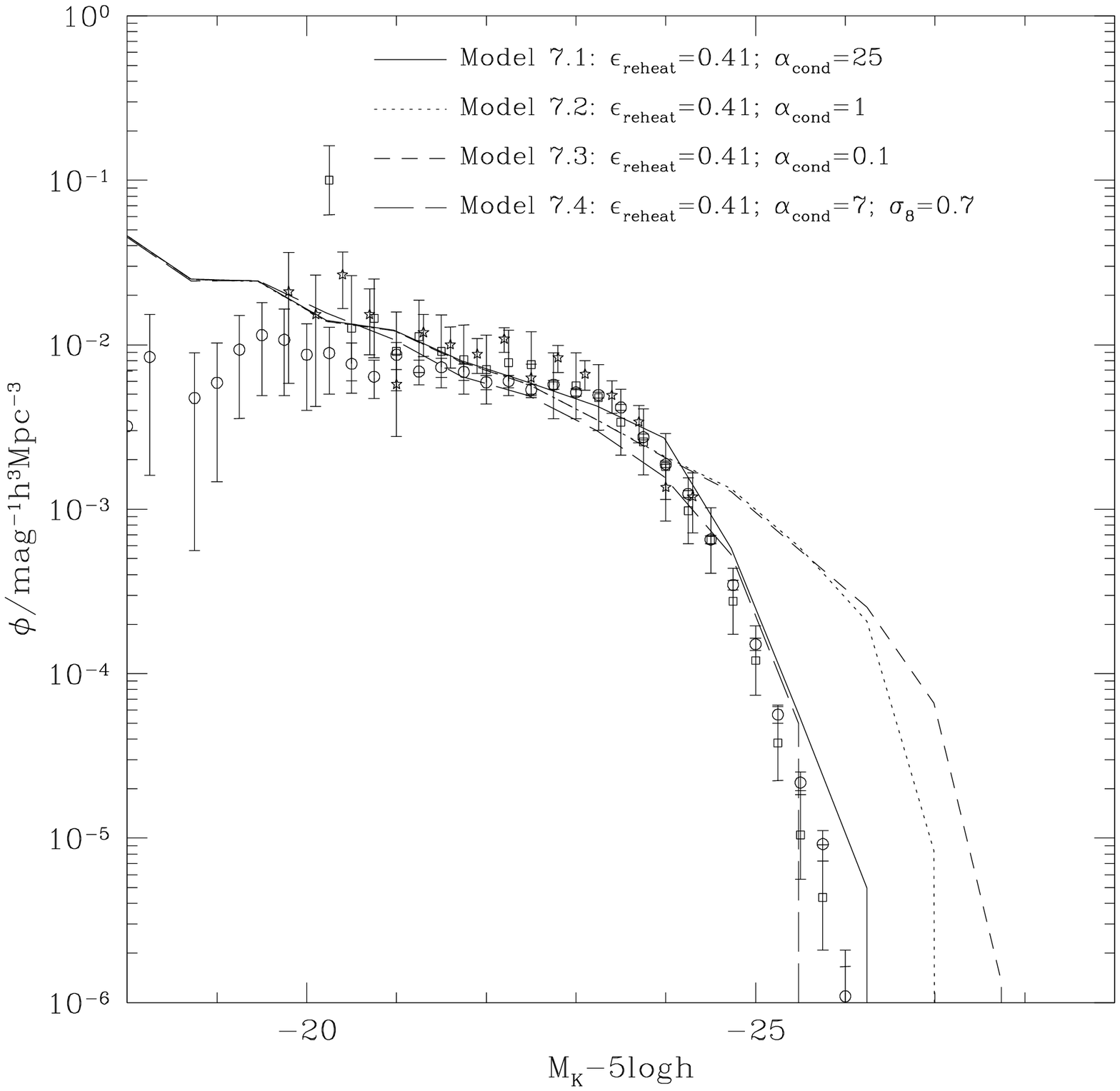}}
\figcaption{Model 7. These models illustrate the effect of thermal
conduction. In Model~7.3 ($\alpha_{\rm cond}=25$), conduction is
assumed to be highly efficient (it is unlikely that such a high
efficiency is physically plausible). More realistic conduction
efficiencies are illustrated in Models 7.2 ($\alpha_{\rm cond}=1$)
and~7.1 ($\alpha_{\rm cond}=0.1$). For Model~7.4, we adopt a lower
value for $\sigma_8$; a conduction efficiency of $\alpha_{\rm cond}=7$
then gives a reasonable match to the observed luminosity function.  In
all cases, the energy feedback parameters parameters are set to
$\eho=0.0$ and $\erh=0.41$.
\label{fig:models_6}}
\end{inlinefigure}

Thermal conduction is expected to imprint a special scale on the
galaxy population because of the strong temperature dependence of the
Spitzer conductivity rate. Model~7.3 (solid line in
Fig.~\ref{fig:models_6}) shows the result of including conduction with
$\alpha_{\rm cond}=25$ in Model~5.3 ($\erh=0.41$, $\eho=0$). Such a
high value of the conductivity is indeed effective in suppressing the
formation of the most massive galaxies since it prevents efficient gas
cooling in group and cluster-sized halos. The brightest galaxies in
the luminosity function are instead built through mergers. The result
is a rather good match to the observed galaxy luminosity
function. However, the conduction efficiency that we have assumed is
extremely high. To operate at the required level, we must assume both
that the conductivity is not suppressed below the Spitzer value, and
that the effective temperature gradient is steeper than $T\propto
r^{2.5}$ in the region of the cooling radius. Note that the Spitzer
formula for thermal conductivity in an ionized plasma breaks down if
the conductivity becomes too high, i.e. if the conduction
``saturates'' as described by \citet{cowie77}. The models shown in
this paper do not take account of this saturation limit. However,
using the estimate of the saturated heat flux from \citet{cowie77}, we
have checked that our results are not significantly affected by
saturation (for model 7.3, which has the most extreme conduction of
all our models, there is only a small increase in the number of the
very brightest galaxies).

In Models 7.1 and 7.2, we show the effect of assuming a more modest
conduction efficiency ($\alpha_{\rm cond}=1.0$ and 0.1
respectively). In these models, conduction is not sufficient to
suppress cooling in the larger halos adequately.

If we adopt a lower value for $\sigma_8$, however, a lower conduction
efficiency gives a reasonable match to the observed luminosity
function.  Model~7.4 shows the luminosity function for the case
$\sigma_8=0.7$ and $\alpha_{\rm cond}=7$. This conduction efficiency
could be achieved if the temperature gradient was $T\propto r^{1.3}$
and the conduction was only slightly suppressed below the Spitzer
value. Although this is still a high rate of conduction, it offers a
promising route for explaining the bright end of the galaxy luminosity
function.

\begin{inlinefigure} 
\resizebox{\textwidth}{!}{\includegraphics{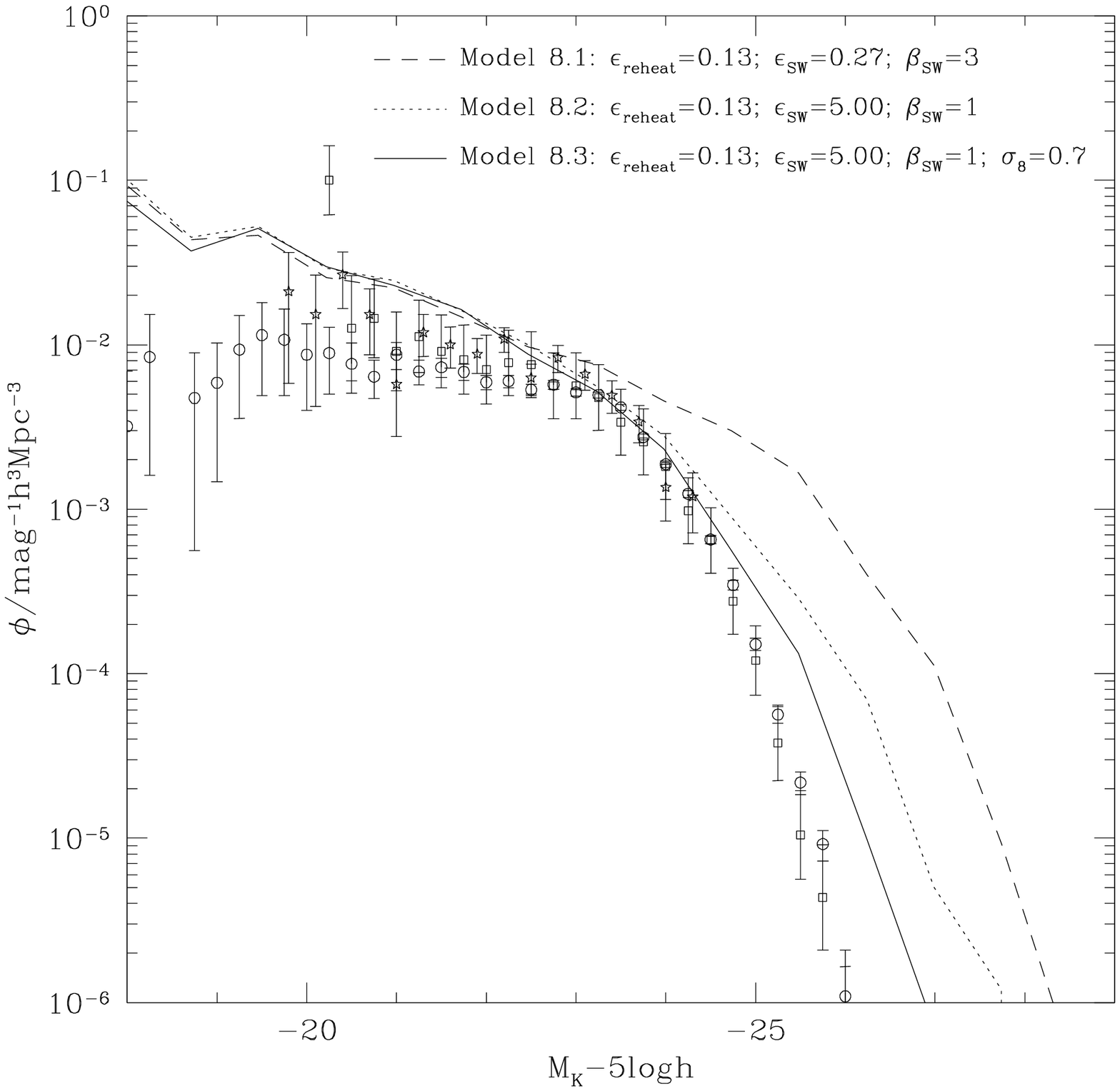}}
\figcaption{Model 8. These models illustrate the effect of superwinds.
In Model~8.1, an energy of $\esw=0.27$ drives a weak superwind (with
$\beta_{\rm SW}=3$); disk reheating has efficiency $\erh=0.13$ and
there is no heating of the diffuse halo ($\eho=0$).  A much more
powerful wind is needed to create a break in the luminosity
function. Model~8.2 ($\esw=5.0$; $\beta_{\rm SW}=1$) illustrates the
effect of increasing the superwind power. An improved match to the
luminosity function can be achieved with the same power if $\sigma_8$
is lower. Model~8.3 shows a model with $\esw=5.0$, $\beta_{\rm SW}=1$
and $\sigma_8=0.7$.
\label{fig:models_7}} 
\end{inlinefigure} 

\subsection{Model with superwinds}

The expulsion of gas from halos at high energy can, in principle,
strongly suppress the formation of  later generations of galaxies,
hence affecting the shape of the luminosity function.  Starting from
Model~5.2, we add further feedback energy that expels cold gas
completely, not only from the disk but also from the halo. The
superwind must have high energy in order that the expelled material
not be recaptured by more massive halos. The effect of a low power
superwind is illustrated by Model~8.1 (dashed line in
Fig.~\ref{fig:models_7}).  This model, with $\esw=0.27$ and
$\beta_{\rm SW}=3$, has a relatively weak superwind. This gas
expulsion is in addition to the reheating of cold disk gas
($\erh=0.13$); we have assumed that there is no heating of the diffuse
halo ($\eho=0.0$). Although the winds eject a large amount of gas,
most of the material is recaptured as larger halos collapse and the
luminosity function differs little from Model~5.2.

In Model~8.2 (dotted line), we have set $\esw=5.0$ and $\beta_{\rm
SW}=1$, corresponding to a mean energy per superwind particle of
$E_{\rm av}=15$keV. Such an energetic wind is required to ensure that
very little material is recaptured by group halos. In this model, the
superwind dominates the feedback energy budget; indeed, the total
energy required ($5.13 \times10^{49}\ergMsol$) significantly exceeds
that available from supernovae alone. The model comes much closer to
matching the luminosity function, but still overproduces bright
galaxies. We can increase the superwind energy further, but a factor
of 2 increase only results in a small improvement in the luminosity
function. If we include conduction as well as superwinds, it is, of
course, possible to improve the match but a high conduction efficiency
($\alpha_{\rm cond}\gg 1$) is still needed.  Increasing the mass
loading of the wind substantially (by increasing $\beta_{\rm SW}$)
results in too few galaxies around the knee of the luminosity function.

As we found in the case of conduction, the luminosity function can be
matched more easily if we adopt a lower value for $\sigma_8$. The case
$\esw=5.0$, $\beta_{\rm SW}=1$, $\sigma_8=0.7$ is illustrated by
Model~8.3 (solid line). Given the uncertainties of our recapture
prescription, this model gives a reasonable match to the luminosity
function; it has a strong break at the correct luminosity, and
overproduces bright objects only marginally. There are a variety of
ways to further improve the match to observations: we could increase
the energy injection (but $\esw > 10.0$ is required) or increase the
mass loading of the wind so that the curve is shifted faintwards.  An
alternative strategy is to allow for a low level of conduction:
$\alpha_{\rm cond}\sim1$ is sufficient to produce a significant
improvement in the match to the luminosity function when
$\sigma_8=0.7$ and superwinds are present.

The parameters $\alpha_{\rm cond}$ and $\epsilon_{\rm SW}$ are highly
degenerate in their effects on the luminosity function
(i.e. increasing either suppresses the bright end). While current
computational limitations make it prohibitively expensive to perform
an accurate $\chi^2$ fit of the model parameters to the data, a crude
estimate of the $\chi^2$ surface for these two parameters shows that
the data prefer models with strong superwinds ($\epsilon_{\rm
SW}\approx 6$) and high conductivity ($\alpha_{\rm cond}\approx 30$)
for a model with $\sigma_8=0.93$. Lowering $\sigma_8$ to $0.7$ reduces
the requirements to $\epsilon_{\rm SW}\approx 3$ and $\alpha_{\rm
cond}\approx 25$. Further investigation of the $\chi^2$ surface would
require a Bayesian prior to specify formally a physically allowed
range for this parameter.

\subsection{Other Considerations}

While our primary goal in this paper is to examine the luminosity
function of galaxies, it is prudent to check whether our models are in
reasonable agreement with other basic properties of the galaxy
population such as the Tully-Fisher relation and the galaxy
autocorrelation function. We compare the models that best fit the
galaxy luminosity function to these observables. We retain the model
parameters found earlier and we do not attempt to adjust these or any
other parameters in this comparison.  We defer a more detailed
comparison of our models with a wider range of observational
constraints to a future paper.

\subsubsection{Tully-Fisher Relation}

Simultaneously matching the galaxy luminosity function and the
Tully-Fisher relation has been a long-standing problem for CDM-based
semi-analytic models (see, for example, \citealt{wf}, \citealt{kwg}
and \citealt{cole00}). In particular, \citet{cole00} found that while
their best model reproduced the observed slope and scatter in the
Tully-Fisher relation, the predicted circular velocities for galaxy
disks were about 30\% larger than the values measured for the data.

Figure~\ref{fig:TF} shows the Tully-Fisher relation predicted by four
of our models (including those which most successfully match the
observed luminosity function). It is readily apparent that our models
all perform reasonably well in that they reproduce the slope and
scatter of the observed relation, although they do not match the
zero-point exactly.  Model~8.3 does best, being offset by only
approximately 10\% in circular velocity over most of the range
shown. This model, in fact, performs rather better than that of
\citet{cole00}, although an even better match would be desirable. This 
may involve revising the assumption of adiabatic invariance made in
calculating the effects of the disk on the inner structure of the dark
matter halo (see \citealt{cole00}) or considering the effects of the
angular momentum of the superwind material or the details of our star
formation prescription. 

\begin{inlinefigure} 
\resizebox{\textwidth}{!}{\includegraphics{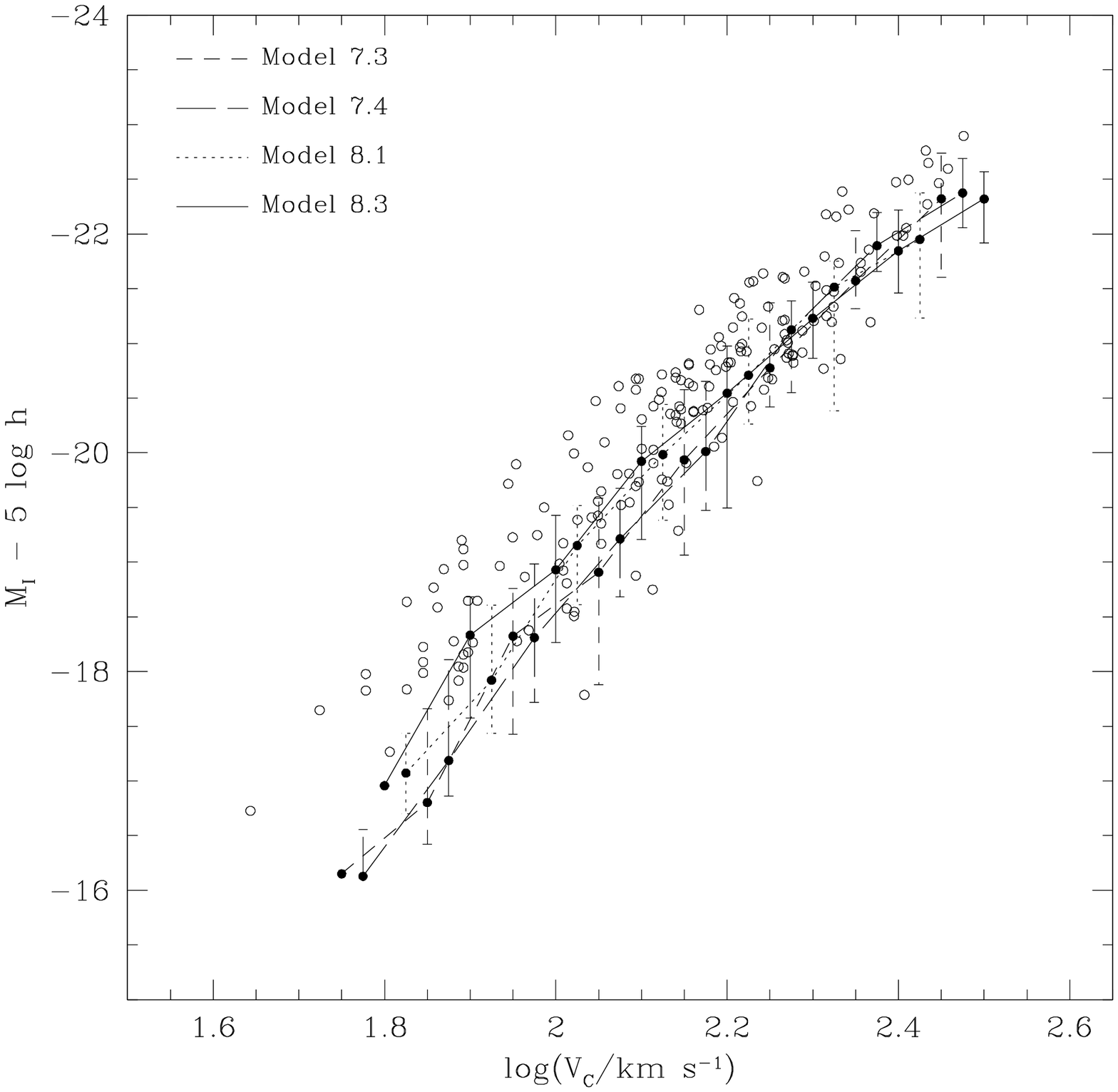}} \figcaption{The
I-band Tully-Fisher relation of galaxies. Circles show the
observational sample of \protect\citet{mathewson}. Lines show the
median relation for model galaxies. We follow \protect\citet{cole00}
and select only galaxies with a bulge-to-total ratio in
dust-extinguished I-band light between $0.02$ and $0.24$ and with a
disk gas fraction of at least 10\%. Circular velocities for model
galaxies are computed at the disk half-mass radius. Errorbars enclose
the inner 80\% of the distribution of model galaxies.
\label{fig:TF}} 
\end{inlinefigure}

\subsubsection{Galaxy Correlation Function}

One of the most remarkable successes of semi-analytic models of galaxy
formation in the $\Lambda$CDM cosmology is their ability to match the
observed galaxy autocorrelation function
(\citealt{kauff99,ajbbias}). As pointed out by \citet{ajbbias}, the
clustering of galaxies may be understood physically by reference to
the halo occupation distribution (see, for example, \citealt{peas01},
\citealt{cooray}), which specifies the number of galaxies that occupy 
dark matter halos of a given mass. We show, in Fig.~\ref{fig:NM}, the
mean number of galaxies brighter than $M_{\rm B}-5\log h=-19.5$ per
halo as a function of halo mass for four of our models. For reference
we also plot the fits obtained by 
\citet{berlind} to the semi-anaytic model of \citet{cole00} and to a
smooth-particle hydrodynamics calculation of galaxy formation as solid
and dotted lines respectively. The two models considered by
\citet{berlind} both gave reasonable fits to the galaxy
correlation function.

\begin{inlinefigure} 
\resizebox{\textwidth}{!}{\includegraphics{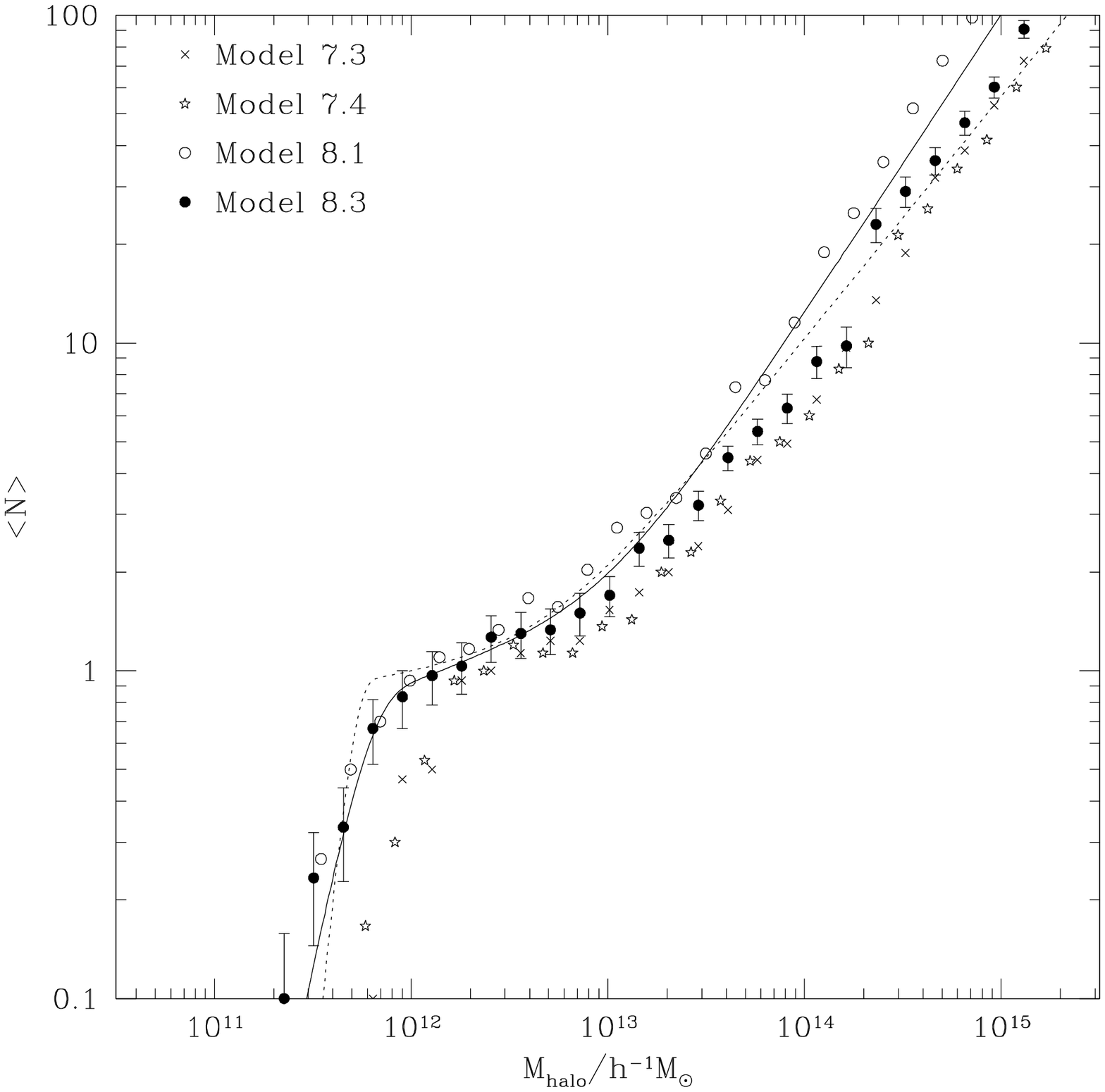}} \figcaption{Symbols
show the mean number of galaxies brighter than $M_{\rm B}-5\log h =
-19.5$ as a function of halo mass for some of our models. For
reference, lines show the fitting functions of
\protect\citet{berlind}. The solid line shows their fit to the results
of the \protect\citet{cole00} semi-analytic model while the dotted
line shows their fit to a smoothed-particle hydrodynamics calculation.
\label{fig:NM}} 
\end{inlinefigure} 

Our new models give rise to a mean halo occupancy which is
qualitatively similar to those found by \citet{berlind} -- a power-law
at high masses, which flattens before cutting off sharply at some
critical mass. Quantitatively, the models with conduction (7.3
and~7.4) produce significantly fewer galaxies in low and high mass
halos, while agreeing with the \citet{berlind} fits at a halo mass of
a few times $10^{12}h^{-1}M_\odot$. The models with superwinds (8.1
and~8.3) agree more closely with the \citet{berlind} fits, although
model 8.3 produces somewhat fewer galaxies in clusters. Since none of
our models differ greatly from those of \citet{berlind}, we may expect
them also to produce correlation functions in reasonable agreement
with the observational data, at least on intermediate and large
scales. On smaller scales, the mean number of galaxies per halo is not
sufficient to specify the correlation function; this requires
knowledge of the distribution of galaxy pairs per halo.

\begin{inlinefigure} 
\resizebox{\textwidth}{!}{\includegraphics{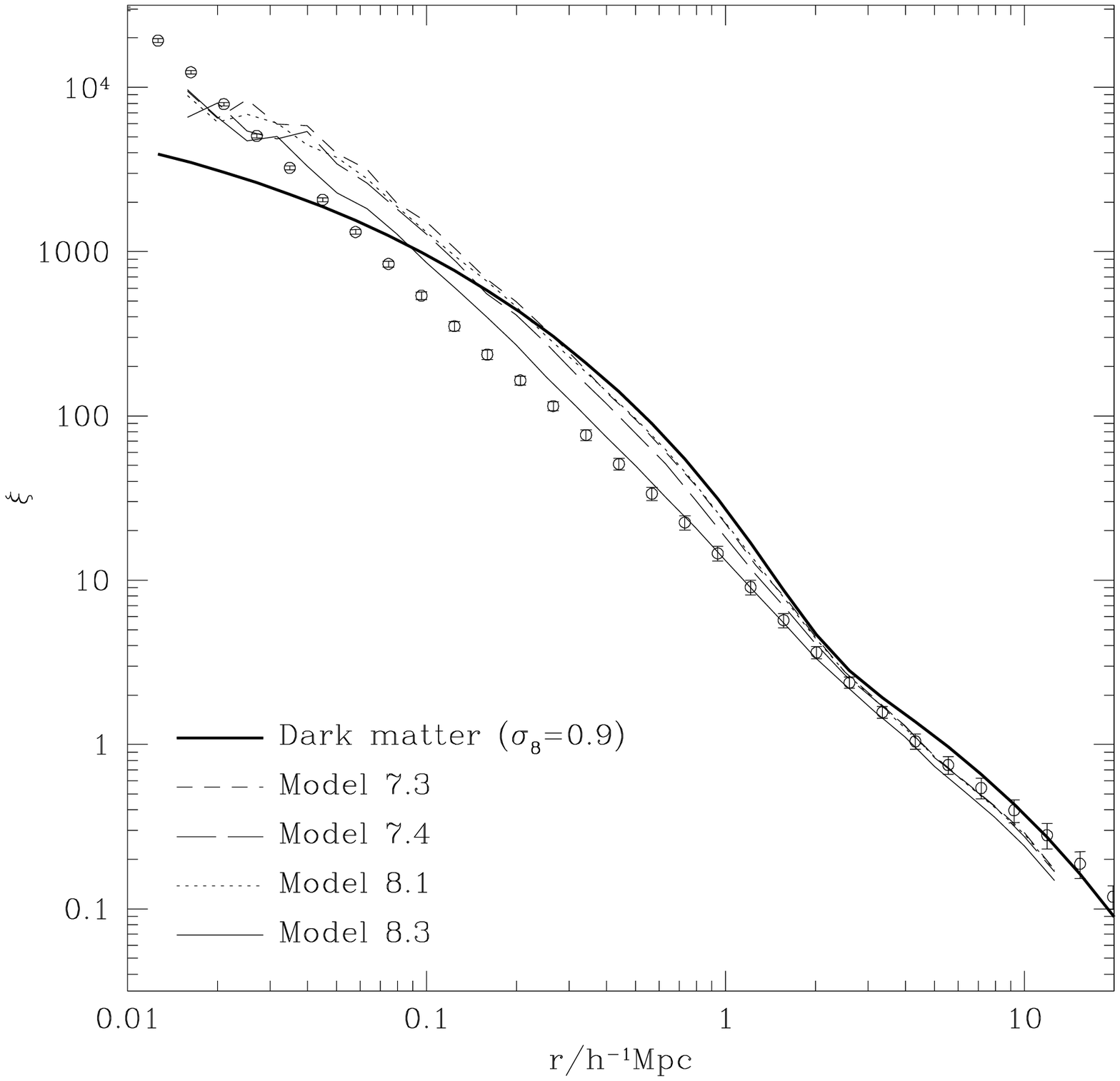}} \figcaption{The
galaxy real-space two-point correlation function. Circles show the
observational determination of \protect\citet{apm}. The heavy solid
line shows the correlation function of dark matter for our adopted
$\Lambda$CDM cosmology (with $\sigma_8=0.9$). The remaining lines show
results for model galaxies brighter than $M_{\rm B}-5\log h=-19.5$, as
indicated in the legend.
\label{fig:xi}} 
\end{inlinefigure} 

Figure~\ref{fig:xi} shows correlation functions (computed using the
techniques of \citealt{ajbbias}) for the same four models, compared to
both the predicted dark matter correlation function and the observed
galaxy correlation function. As expected, all models do well at
matching the data on scales larger than about $1h^{-1}$Mpc. On smaller
scales, all models are `antibiassed' with respect to the dark matter,
but they produce somewhat too much clustering. \citet{ajbbias} found
similar results for their semi-analytic model. Model~8.3 gives, in
fact, very similar clustering results to theirs (for the same
$\Lambda$CDM cosmology). The other models produce even stronger
correlations on small scales, providing a worse match to the
data. Model~8.3 does better than Model~8.2 because it places
relatively fewer galaxies in clusters, thereby having many fewer pairs
on small scales. We speculate that a model which performs even better
at matching the observed luminosity function (e.g. producing a sharper
cut-off at bright magnitudes) may also do even better at matching the
observed correlation function.

\section{Discussion}

We have examined the key physics thought to be necessary to explain
the shape of the galaxy luminosity function in a cold dark matter
universe. While the cooling and condensation of gas in a merging
hierarchy of dark matter halos remains the fundamental process through
which galaxies form, we have demonstrated that at least two other
processes must act to shape the luminosity function. A \emph{minimal}
model requires the inclusion of feedback mechanism(s) (beyond the
heating resulting from the photoionization of the pregalactic gas) to
flatten the faint end of the luminosity function and to suppress
cooling at the centres of the massive halos of groups and clusters. If
a fraction of the energy liberated by supernovae and winds goes into
reheating disk gas and/or heating the diffuse gas halo, then the
formation of faint galaxies is suppressed, resulting in a flattened
faint-end slope that matches the available observational data
adequately. These mechanisms on their own, however, are unable to
produce a sharp cut-off at the bright end of the luminosity
function. We have shown that there are two possible (but quite
extreme) processes that can achieve this: (1) thermal conduction at
about or above the Spitzer rate; (2) expulsion of gas from halos in
superwinds at temperatures high enough to prevent its subsequent
recapture.

The high value of $\alpha_{\rm cond}$ required to suppress the bright
end of the luminosity function is discouraging since it implies both
that the conductivity must be close to the Spitzer value (despite the
presence of $\mu G$ magnetic fields in clusters;
\citealt{narayan01,taylor02}), and that the effective temperature
gradient must be somewhat steeper than the gradients observed in galaxy
clusters \citep{allen01, fabian02}.

However, our method for calculating the effect of conduction is highly
simplified.  In particular, there are two issues that are not well
addressed. Firstly, our calculation is based on the heat flux through
a shell. We have not considered the total extent of the region in
which conduction must be effective. We can define a radius $r_{\rm
out}$, such that the initial thermal energy in the region $r_{\rm
cool} <r< r_{\rm out}$ is equal to the total energy radiated from the
region $r< r_{\rm cool}$, where $r_{\rm cool}$ is the cooling radius
in the presence of conduction. Thermal energy needs to be conducted
from a radius at least as large as $r_{\rm out}$ in order to balance
the radiative losses from smaller radii.  For an isothermal halo with
an $r^{-2}$ density profile, we find $r_{\rm cool} r_{\rm out} =
r^{\prime 2}_{\rm cool}$, where $r^\prime_{\rm cool}$ is the cooling
radius calculated in the absence of conduction. The radius $r_{\rm
out}$ can be large, but the temperature gradient that is required to
transport heat effectively across this region is a strongly declining
function of radius. Thus, so long as $r_{\rm out}$ lies within the
virialized region of the cluster, this is unlikely to modify our
solution significantly. Secondly, our method assumes that the
logarithmic temperature gradient within $r_{\rm cool}$ is
constant. If, instead, the gradient is peaked in this region, our
method would underestimate the heat deposited in the shell at $r_{\rm
cool}$. These two simplifications (which affect the required
conduction coefficient in opposite senses) are difficult to model
accurately and without ad-hoc assumptions. Clearly, this is a problem
that needs to be addressed by numerical simulations of cooling in
conductive plasmas. These would allow us to calibrate our simple model
of conductive heating and to infer what conductivity is required, by
removing the degeneracy between the shape of the halo temperature
profile and the suppression factor $f_{\rm sp}$ for the conductivity.

The alternative model invokes highly energetic ``superwind'' outflows.
The expulsion of gas from the halo reduces the reservoir of baryons
available for cooling at later times, and reduces the number of stars
formed by each parcel of cold gas. However, the fraction of baryons
required to be in this ``super-heated'' phase at $z=0$ is quite high,
and the energy budget is exceptional. We found the energy budget of a
promising superwind model to be in region of $5 \times 10^{49}$ergs
per Solar mass of stars formed.  It is quite possible our model
may not treat the recapture process adequately and that we may
therefore be overestimating the total energy requirement. It is
difficult to model the process better, however, because the expelled
gas may cool adiabatically and then be accreted by a completely
different halo; alternatively, some material may escape into void
regions and evade recapture altogether. A further problem is the need
to assume a particular distribution for the wind particle
energies. Our assumption that the distribution is thermal may lead us
to overestimate the energy requirement.

On average, the mass of baryons ejected from halos amounts to 7\% of
mean cosmic value. This corresponds to about 1.3 times the fraction of
baryons turned into stars. A compilation of observations by
\cite{martin99} suggests that the total mass loss rate from disk
galaxies could certainly be as high as twice the star formation
rate. If all of this mass loss were energetic enough to produce a
superwind, these observations would provide some justification for the
high value of $\esw$ required in our model. The total energy budget,
however, exceeds that available from supernovae ($\sim 0.9 \times
10^{49} \,\hbox{erg}\,M_{\odot}^{-1}$ for the Kennicutt IMF) even
though we have not allowed for radiative losses. Although the exact
energy output from supernovae is uncertain, and the conditions we have
assumed for superwind escape are also uncertain, this suggests that
the supernova energy must need to be supplemented by energy released
during black hole formation
\citep{ensslin98, bower01, cavaliere02}.

Finally, we have briefly examined the ability of some of our most
successful models to match the observed Tully-Fisher relation and
galaxy correlation function. Without adjusting any parameters, we
find, as in many previous semi-analytic calculations, that our models
adequately match the slope and scatter of the Tully-Fisher relation,
but tend to overpredict the circular velocity at a given luminosity
(or, equivalently, underpredict the luminosity at given circular
velocity.) In most cases, the agreement is at least as good as in
previous studies, but for one of our superwind models the discrepancy
in the circular velocity zero-point is reduced to only 10\%.  At least
one of our models (Model~8.3) matches the galaxy autocorrelation
function reasonably well (as well as the semi-analytic model of
\citet{ajbbias}). Importantly, this success reflects the fact that
clusters in this model contain relatively fewer galaxies, and thus
many fewer pairs on small scales, than clusters in other models.  This
demonstrates how measurements of galaxy clustering place important
constraints on halo mass-dependent feedback processes such as those
considered in this study.  Our results encourage a more detailed
investigation of how our models fare when compared with a wider range
of observables, a programme that we intend to pursue in future work.

\section{Conclusions}

One of the simplest properties of the galaxy population, the
luminosity function, is now reasonably well determined observationally
(although, as we have noted, some significant disagreements about the
faint-end slope remain). We have shown that our theoretical
understanding of this fundamental property has progressed enormously
in recent years to the extent that it is now possible to explain the
form of the luminosity function in quantitative detail. Nevertheless,
two crucial ingredients---feedback and conduction---are understood
only poorly at best. Further study of these physical processes is
required before we can truly claim a satisfactory understanding of the
luminosity function of galaxies.

\section*{Acknowledgments}

We thank J.-S. Huang and collaborators for providing their K-band
luminosity function data in electronic form and Simon White and Peder
Norberg for helpful conversations. AJB acknowledges the hospitality of
the Institute of Computational Cosmology of the University of Durham
where much of this work was completed. RGB acknowledges support from a
Leverhulme Fellowship. This work was supported in part by the UK
Particle Physics and Astronomy Research Council, the Royal Society and
the EC Research Network for Research into the Physics of the
Intergalactic Medium.

\appendix

\section{Ejection and Recapture of Superwinds}
\label{app:fracs}

The wind is ejected at speed $V_{\rm SW}$, so has mean energy per unit
mass of $E_{\rm av} = \frac{1}{2}V_{\rm SW}^2$. We assume that to
escape the halo it needs energy per unit mass $\lambda_{\rm SW}V_{\rm
disk}^2$. Assuming a thermal distribution of energies in the wind
(i.e. $f(E) \propto \exp(-E/E_{\rm av})$), the fraction of the mass
flowing out of the disk which also escapes from the halo is given by
\begin{equation}
f_{\rm eject} = \int_{x_{\rm SW}}^\infty \exp(-x) \d x = \exp(-x_{\rm
SW}),
\end{equation}
where $x = E/E_{\rm av}$ and $x_{\rm SW} = \lambda_{\rm SW} V_{\rm
disk}^2/E_{\rm av}$. For the material which escapes the halo, the mean
energy excess over that needed to eject it from the halo is
\begin{equation}
E_{\rm esc} = \left. E_{\rm av} \int_{x_{\rm SW}}^\infty (x-x_{\rm
SW}) \exp(-x) \d x \right/ \int_{x_{\rm SW}}^\infty \exp(-x) \d x =
E_{\rm av},
\end{equation}
i.e. the mean (kinetic+thermal) energy per unit mass for the gas escaping from
the halo is the same as the mean (kinetic+thermal) energy per unit
mass for the  gas flowing out of the disk (since only the highest
energy particles escape). Using therefore 
the same energy distribution, we assume that the fraction of particles
recaptured in a halo of virial velocity $V_{\rm halo}$ is
\begin{equation}
f_{\rm cap} = \int^{x_{\rm cap}}_0 \exp(-x) \d x = 1 - \exp(-x_{\rm cap})
\end{equation}
where $x_{\rm cap} = C_{\rm cap} V_{\rm halo}^2/E_{\rm av}$, and
$C_{\rm cap}$ is a numerical coefficient of order unity. We have used
$C_{\rm cap}=1$. 

\end{document}